\documentclass[conference, 10pt]{IEEEtran}
\IEEEoverridecommandlockouts

\usepackage{cite}
\usepackage{amsmath,amssymb,amsfonts}
\usepackage{algorithmic}
\usepackage{graphicx}
\usepackage{textcomp}
\usepackage{booktabs}
\usepackage{siunitx}
\usepackage{multirow}
\usepackage{makecell}
\usepackage{bm}
\usepackage[table,xcdraw]{xcolor}
\def\BibTeX{{\rm B\kern-.05em{\sc i\kern-.025em b}\kern-.08em
    T\kern-.1667em\lower.7ex\hbox{E}\kern-.125emX}}
\begin{document}
\newcommand{\grayvline}{\tikz[baseline]{\draw[lightgray, line width=1pt] (0,0em) -- (0,0.6em);}}
\sisetup{
  detect-weight,
  detect-inline-weight=math,
  table-format = 2.2
}
\title{Multi-Distillation from Speech and Music Representation Models}

\author{\IEEEauthorblockN{Jui-Chiang Wei$^{*}$}
\IEEEauthorblockA{\textit{National Taiwan University} \\
Taiwan \\
johnwei743251@gmail.com}
\and
\IEEEauthorblockN{Yi-Cheng Lin$^{*}$}
\IEEEauthorblockA{\textit{National Taiwan University} \\
Taiwan \\
f12942075@ntu.edu.tw}
\and

\IEEEauthorblockN{Fabian Ritter-Gutierrez}
\IEEEauthorblockA{\textit{Nanyang Technological University} \\
Singapore \\
s220064@e.ntu.edu.sg}
\and
\IEEEauthorblockN{Hung-yi Lee}
\IEEEauthorblockA{\textit{National Taiwan University} \\
Taiwan \\
tlkagkb93901106@gmail.com}
\and
}

\maketitle

\begingroup
  \renewcommand\thefootnote{*}
  \footnotetext{Equal contribution}
\endgroup

\begin{abstract}

Real-world audio often mixes speech and music, yet models typically handle only one domain. This paper introduces a multi-teacher distillation framework that unifies speech and music models into a single one while significantly reducing model size. Our approach leverages the strengths of domain-specific teacher models, such as HuBERT for speech and MERT for music, and explores various strategies to balance both domains. Experiments across diverse tasks demonstrate that our model matches the performance of domain-specific models, showing the effectiveness of cross-domain distillation. Additionally, we conduct few-shot learning experiments, highlighting the need for general models in real-world scenarios where labeled data is limited. Our results show that our model not only performs on par with specialized models but also outperforms them in few-shot scenarios, proving that a cross-domain approach is essential and effective for diverse tasks with limited data.

\end{abstract}

\begin{IEEEkeywords}
self-supervised models, knowledge distillation
\end{IEEEkeywords}

\section{Introduction}
Self-supervised learning (SSL) has led to significant advancements in both speech \cite{mohamed2022self, superb} and music processing \cite{marble}. Models like HuBERT \cite{hubert} and WavLM \cite{wavlm} have demonstrated exceptional performance across various speech tasks, while MERT \cite{mert} has shown similar success in music understanding tasks. 
Although speech and music share auditory features like pitch, rhythm, and timbre, they follow different patterns and pose unique representation challenges. Speech models are primarily designed to capture linguistic content, prosody, and phonetic nuances \cite{wav2vec2, TERA}. In contrast, music models are optimized for melodic structure, harmonic progression, and timbral texture \cite{huang2024dynamic, jukebox, CLMR}. As a result, applying a speech model to music-related tasks often leads to misinterpretation of musical features as noise or irrelevant variations. Conversely, music models may overlook the fine-grained temporal and semantic details critical for speech understanding.

In real-world settings such as podcasts with background music or multimedia content that combines dialogue and soundtracks, audio signals rarely conform to a single domain \cite{EnCodecMAE, rittergutierrez2025distillingspeechmusicencoder}.
This issue has been recognized in recent works on universal audio representation learning \cite{universal}, which show that speech and music models struggle to generalize beyond their trained domains. This disconnection presents a critical challenge: \emph{how can we build a unified model that robustly interprets both speech and music?} 

        
    Motivated by this need, we adopt a multi-teacher distillation framework that merges the strengths of domain-specific teacher models into a single student model. Building on prior research in multi-teacher and ensemble distillation \cite{EKD, theia} and adaptive loss weighting techniques \cite{EAKD}, we conduct a series of ablation studies to evaluate the impact of several domain-adaptive strategies on cross-domain performance.
    
    First, we explore Data-Domain Separation (DDS), in which speech and music data are routed to their respective teacher models during training. This approach aims to preserve each domain's unique characteristics while minimizing cross-domain interference. Second, we investigate various translation modules: linear, convolutional, and hybrid, to align the diverse representations of the teacher models with those of the student model, preserving critical domain-specific details. Finally, we adjust each teacher's loss to balance the influence of the speech and music models, preventing one domain from dominating the learning process.

    Our ablation studies demonstrate that the resulting unified student model performs competitively on speech and music tasks, achieving high accuracy while reducing the computational burden of maintaining separate models. This cross-domain capability opens new avenues for applications such as speech-based music recommendation, audio content indexing, and intelligent multimedia analysis.

    Additionally, we conduct few-shot learning experiments to evaluate how well our unified model generalizes under low-resource conditions. In real-world applications, labeled data is often scarce, and deploying task-specific models for every domain is impractical. Instead, what is truly needed is a single, general model that can handle diverse tasks and still adapt effectively with limited supervision. The results show that our unified model consistently outperforms its domain-specific counterparts in few-shot settings, demonstrating that cross-domain distillation not only yields a more versatile model but also one that is better suited for real-world, resource-constrained environments.

    In summary, our contributions are threefold:
    \begin{itemize}
       \item We identify and address the underexplored problem of cross-domain audio modeling, focusing on unifying speech and music representations within a single student model using multi-teacher distillation.
        \item We systematically evaluate domain-adaptive strategies, including data separation, feature translators, and loss weighting and analyze their effect on multi-domain distillation performance.
        \item We validate our approach through comprehensive experiments across diverse downstream tasks, including few-shot learning, highlighting its effectiveness and potential impact in real-world audio processing scenarios.
    \end{itemize}

\begin{figure*}[htbp]
  \centering
  \includegraphics[width=1.0\textwidth]{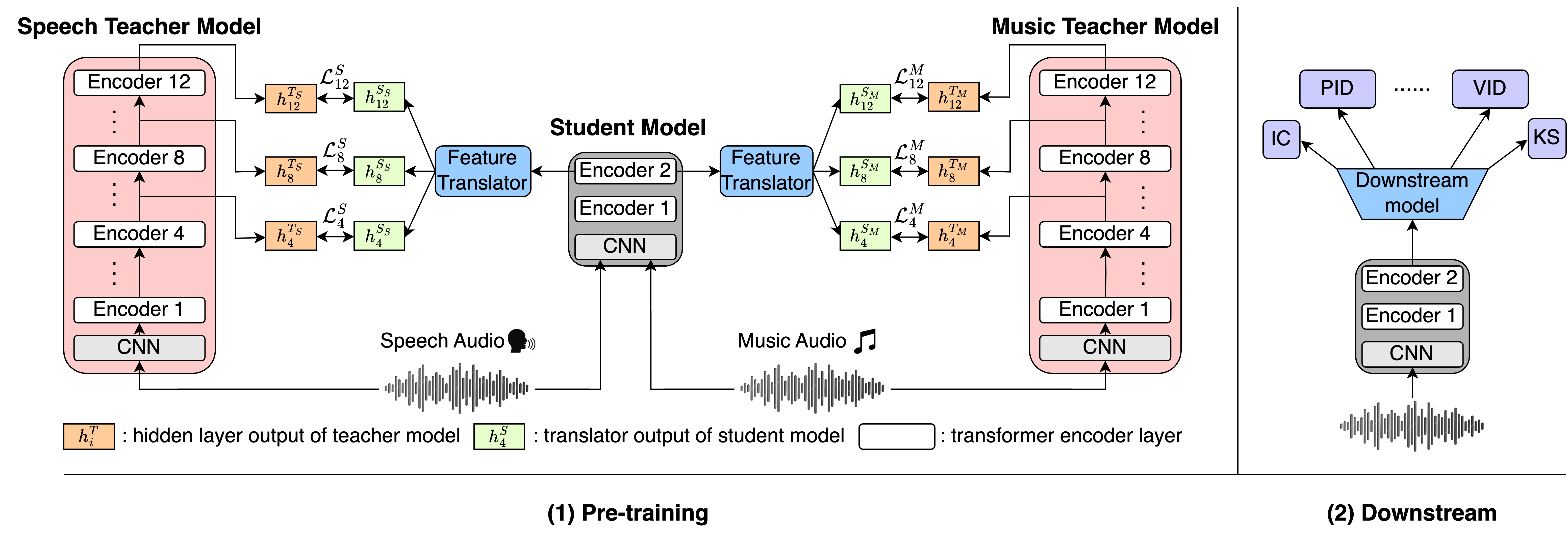}
  \caption{Illustration of multi-distillation structure}
  \label{fig:myfigure}
\end{figure*}

\section{Related Work}
\subsection{Knowledge Distillation}

Knowledge distillation (KD) \cite{jinyu_distill, hintonDistill} is a widely used technique for compressing large teacher models into smaller, efficient student models while retaining much of the teacher's performance. This approach has been applied extensively in self-supervised learning (SSL). For example, DistilHuBERT \cite{distilhubert} successfully distilled HuBERT into a smaller model using prediction heads to align student and teacher representations. FitHuBERT and LightHuBERT \cite{fithubert, lighthubert} extended this concept through parameter-efficient fine-tuning and architectural modifications, maintaining performance with significantly reduced resource requirements. Prior works have also explored distillation for debiasing SSL models \cite{lin24b_interspeech} or noise-robust SSL models \cite{robustdistiller,ritter2023noise, RobustDistilHuBERT}.

Similarly, in the music domain, KD has been used to accelerate large models for music generation and classification; for instance, Fast Jukebox \cite{pezzat2023fast}, singing voice detection \cite{paul21b_interspeech} and symbolic instrument classification \cite{instrument_distill} show that student networks inherit rich musical expertise from the teachers.

\subsection{Multi-Teacher and Ensemble Distillation}
Multi-teacher learning extends knowledge distillation to aggregate knowledge from multiple sources. 
Studies have shown that multi-teacher distillation can combine complementary expertise and often yields a student model that outperforms those distilled from any single teacher \cite{zuchniak2023multi}. 
In natural language tasks, multi-teacher frameworks have been explored for multi-domain and multi-task scenarios: \cite{ilichev-etal-2021-multiple} distilled knowledge from teachers trained on different tasks.
\cite{tan2018multilingual} similarly used a separate teacher for each language pair in machine translation to build a single multilingual student. 
In computer vision, Theia \cite{theia} explores multi-model distillation in vision by using feature translators to merge diverse vision foundation models. These studies show that a student network can integrate a wide range of knowledge, from different task objectives to varied feature types, through coordinated distillation.
However, none of these works discuss how to manage the unique characteristics and complexities of speech data.

In the audio domain, ensemble-based approaches have been explored to combine multiple supervised models \cite{Wu2021OneTI, Wu2022UnifiedAE, Gao2020DistillingKF, Chebotar2016DistillingKF}. EKD \cite{EKD} distills multiple self-supervised speech models into a unified student model using layerwise averaging and task-specific prediction heads.
\cite{yang2023knowledge} also proposed distilling multiple foundation models into one ASR system. 
While effective for speech tasks like ASR and emotion recognition, these works remain limited to the same-domain distillation. 
Our work extends this to cross-domain scenarios, integrating speech and music models while adopting extensive methods. 

\begin{table*}[t]
  \centering
  \caption{Downstream tasks, metrics, datasets, and domains for speech and music tasks. The symbol $\downarrow$ indicates that a lower score is better, while $\uparrow$ indicates that a higher score is better.}
  
  \setlength{\tabcolsep}{6pt}
  \renewcommand{\arraystretch}{1.15}
  \begin{tabular}{ l l c l }
    \toprule
    \textbf{Downstream Tasks} & \textbf{Evaluation Metrics} & \textbf{Dataset} & \textbf{Domain} \\
    \midrule
    \textbf{ASR} (Auto. Speech Recognition) & word error rate (WER) $\downarrow$ & LibriSpeech \cite{librispeech} & Speech \\
    \textbf{KS} (Keyword Spotting) & accuracy (ACC) $\uparrow$ & Fluent Speech Commands \cite{fsc} & Speech \\
    \textbf{IC} (Intent Classification) & accuracy (ACC) $\uparrow$ & Speech Commands \cite{speech_commands} & Speech \\
    \textbf{SID} (Speaker Identification) & accuracy (ACC) $\uparrow$ & VoxCeleb1 \cite{voxceleb} & Speech \\
    \textbf{ER} (Emotion Recognition) & accuracy (ACC) $\uparrow$ & IEMOCAP \cite{iemocap} & Speech \\
    \textbf{SingID} (Singer Identification) & accuracy (ACC) $\uparrow$ & VocalSet \cite{vocalset} & Music \\
    \textbf{VID} (Vocal technique Identification) & accuracy (ACC) $\uparrow$ & VocalSet \cite{vocalset} & Music \\
    \textbf{PID} (Pitch Identification) & accuracy (ACC) $\uparrow$ & NSynth \cite{nsynth} & Music \\
    \bottomrule
  \end{tabular}
\label{tab:downstream_tasks_metrics}
\end{table*}

\section{Multi-Teacher Distillation}
We propose a multi-teacher distillation framework that fuses complementary knowledge from speech and music domains into a single student model. The framework is depicted in Figure ~\ref{fig:myfigure}. Two pretrained teacher models: one for speech (e.g., HuBERT or WavLM) and one for music (e.g., MERT), provide domain-specific representations.

During training, input data is first separated by domain: speech samples are processed by the speech teacher, and music samples by the music teacher, ensuring that each domain's specialized features are captured independently. To align the student model's representation with those of the teachers, a feature translator (e.g., a linear or convolutional module) is applied to the student model's output, transforming it into a space that matches the teacher models' representations.

Once the student model's representations are transformed, learning is guided by a layerwise distillation objective. For each selected layer $l$ and each teacher $m$ (speech or music), the per-layer loss is given by:
\begin{equation}
    \mathcal{L}_{l,m} = \frac{1}{D^T} ||h_l^{S_m} - h_l^{T_m}||_1 - \log \sigma\Bigl(\text{cossim}\bigl(h_l^{S_m}, h_l^{T_m}\bigr)\Bigr),
\end{equation}
where $h_l^{S_m}$ and $h_l^{T_m}$ denote the student and teacher features, $D^T$ is the teacher feature dimension, and $\sigma$ is the sigmoid function. This dual loss function is adopted from DistilHuBERT \cite{distilhubert}, which combines L1 loss for magnitude alignment and cosine similarity loss for directional consistency. This approach preserves both the scale and semantic structure of the representations. As it has proven effective in prior work, we do not explore it further.

The overall loss for music and speech is then computed as:
\begin{equation}
\mathcal{L}_{\text{multi-distill}} =  \sum_{l \in \{4,8,12\}} \sum_{m \{\text{S}, \text{M}\}} w_m\mathcal{L}_{l,m}\bigl(h_l^{S_m}, h_l^{T_m}\bigr),
\end{equation}
where $w_m$ controls the contribution of each teacher, ensuring that neither domain disproportionately dominates the learning process. This framework enables the student model to inherit specialized knowledge from both domains while harmonizing conflicting representations, ultimately yielding a unified audio representation that performs robustly across diverse tasks.

\section{Experiment Setup}\label{sec:experiments}
\subsection{Pretraining and Data Sampling}

We pretrain the student model within our multi-teacher distillation framework on 960 hours of LibriSpeech \cite{librispeech} for speech and 1,000 hours of Music4ALL \cite{music4all} for music. Each training batch (batch size 24) is formed by randomly sampling from both datasets. The student model consists of a CNN feature extractor and a 2-layer transformer encoder, initialized with the first layers of the speech teacher. Training is conducted for 200k updates using the Adam \cite{adam} optimizer with a learning rate of 2e-4.

We use two pretrained teacher models: a speech teacher (HuBERT or WavLM) and a music teacher (MERT) for the distillation. Both teachers use a 12-layer transformer encoder, and we select intermediate layers (e.g., 4, 8, and 12) as distillation targets.

\begin{table*}[t!]
  \centering
  \caption{Overall performance on speech and music tasks. HB, WL, and MR denote HuBERT, WavLM, and MERT. Multi-distillation models adopt all experimental strategies, with "+" indicating the two teacher models (e.g., HB + MR distills HuBERT and MERT). All multi-distilled models are initialized from their speech teacher. Bold numbers indicate the best-average score.}
  \label{tab:overall}
  \setlength{\tabcolsep}{6pt}
  \renewcommand{\arraystretch}{1.15}
  \begin{tabular}{ ll *{9}{c}}
    \toprule
        \textbf{Model} & \textbf{\#params} 
      & \textbf{ASR} & \textbf{KS} & \textbf{IC} & \textbf{ER} 
      & \textbf{SID} & \textbf{SingID} & \textbf{VID} & \textbf{PID} & \cellcolor{gray!20} \textbf{Avg (w/o ASR)} \\
          \cmidrule(lr){3-3} \cmidrule(lr){4-11} \vspace{-10pt}\\
                     &  (M)             
      & \textbf{(WER\% $\downarrow$)} 
      & \multicolumn{8}{c}{\textbf{(Acc\% $\uparrow$)}} \\
    \midrule
    distill-HuBERT   & 23   & 13.18 & 95.36 & 95.04 & 64.88 & 71.52 & 80.50 & 63.33 & 21.00 & \cellcolor{gray!20} 70.23 \\
    distill-WavLM    & 23   & 13.79 & 95.59 & 94.67 & 60.55 & 63.30 & 78.65 & 61.84 & 51.00 & \cellcolor{gray!20} 72.23 \\
    distill-MERT     & 23   & 23.21 & 86.89 & 52.39 & 59.45 & 33.37 & 85.62 & 69.12 & 89.87 & \cellcolor{gray!20} 68.10 \\
    multi-distill\_HB+MR & 23 & 14.65 & 95.26 & 93.36 & 61.85 & 69.29 & 84.27 & 71.23 & 87.55 & \cellcolor{gray!20} \textbf{80.40} \\
    multi-distill\_WL+MR & 23 & 13.98 & 95.49 & 94.52 & 59.88 & 59.77 & 81.78 & 70.53 & 85.25 & \cellcolor{gray!20} 78.17 \\
    \midrule
    distill-HB/MR\_Ensemble & 46 & 14.65 & 94.26 & 92.59 & 62.78 & 69.88 & 90.25 & 71.23 & 89.28 & \cellcolor{gray!20} 81.16 \\
    distill-WL/MR\_Ensemble & 46 & 14.33 & 96.04 & 93.09 & 62.19 & 65.24 & 90.04 & 71.23 & 89.23 & \cellcolor{gray!20} 80.48 \\
    HuBERT (HB) & 95   & 6.42  & 96.30 & 98.34 & 64.92 & 81.42 & 81.49 & 65.35 & 70.04 & \cellcolor{gray!20} 79.69 \\
    MERT (MR)   & 95   & 23.61 & 89.45 & 59.21 & 55.19 & 29.49 & 85.69 & 72.54 & 91.26 & \cellcolor{gray!20} 68.98 \\
    WavLM base+ (WL) & 95 & 5.59 & 97.37 & 99.00 & 68.65 & 89.00 & 84.27 & 74.34 & 61.74 & \cellcolor{gray!20} \textbf{82.05} \\
    \bottomrule
  \end{tabular}
\end{table*}

\begin{table*}[t]
  \centering
  \caption{Performance of Data-Domain separation (DDS). w/o DDS for not distinguishing the domain, Libri for LibriSpeech. Compares the performances for HB/WL + MR separately.}
  \label{tab:data_domain} 
  \setlength{\tabcolsep}{6pt}
  \renewcommand{\arraystretch}{1.15}
  \begin{tabular}{ ll *{9}{c}}
    \toprule
        \textbf{Teacher} & \textbf{Data} 
      & \textbf{ASR} & \textbf{KS} & \textbf{IC} & \textbf{ER} 
      & \textbf{SID} & \textbf{SingID} & \textbf{VID} & \textbf{PID} & \textbf{\cellcolor{gray!20}Avg (w/o ASR)} \\
          \cmidrule(lr){3-3} \cmidrule(lr){4-11} \vspace{-10pt}\\
                     &               
      & \textbf{(WER\% $\downarrow$)} 
      & \multicolumn{8}{c}{\textbf{(Acc\% $\uparrow$)}} \\
    \midrule
    HB + MR & DDS  &  15.31  & 94.19        & 93.00 & 60.72       & 58.67   & 84.56           & 70.52        & 89.36        & \cellcolor{gray!20} \textbf{78.71} \\
    HB + MR & w/o DDS  & 15.75 & 94.94 & 86.40         & 61.21       & 59.91   & 87.33   & 71.14 & 89.85 & \cellcolor{gray!20} 78.68      \\
    HB + MR & Libri& 15.04 & 94.29        & 85.05         & 61.49 & 61.78 & 82.00           & 70.00        & 86.01        & \cellcolor{gray!20} 77.23      \\
    \midrule
    WL + MR & DDS  & 15.40 & 94.64        & 87.82 & 61.19 & 56.90 & 86.90           & 68.16        & 89.14        & \cellcolor{gray!20} \textbf{77.69} \\
    WL + MR & w/o DDS  & 15.51 & 94.87 & 83.55         & 61.19       & 55.10   & 87.26   & 69.12 & 89.97 & \cellcolor{gray!20} 77.29      \\
    WL + MR & Libri& 15.05 & 94.97        & 87.08         & 60.38       & 56.90 & 84.91           & 68.18        & 86.47        & \cellcolor{gray!20} 76.99      \\
    \bottomrule
  \end{tabular}
\end{table*}

\subsection{Teacher Models}

We employ three SSL teacher models in our work: HuBERT and WavLM as speech models, and MERT as a music model:

\begin{itemize}
    \item \textbf{HuBERT} \textbf {(HB)} is a self-supervised speech model that learns representations by predicting discretized cluster assignments from masked audio inputs. In our experiments, we adopt the base variant\footnote{https://huggingface.co/facebook/hubert-base-ls960}. 
    \item \textbf{WavLM} \textbf {(WL)} is an improved variant of HuBERT that enhances speech representation by incorporating gated relative position bias and data augmentation strategies. It is pretrained on 94k hours of diverse speech data. 
    In our experiments, we use the base+ variant of WavLM\footnote{https://huggingface.co/microsoft/wavlm-base-plus}.
    \item \textbf{MERT} \textbf {(MR)} 
    is a self-supervised music representation model. 
    It leverages transformer encoders to capture music-specific characteristics such as melody, timbre, and rhythm. Our experiments use the variant pretrained on the Music4ALL dataset (mert-public-v0) \footnote{https://huggingface.co/m-a-p/MERT-v0-public}.
\end{itemize}

\subsection{Downstream Tasks Evaluation}

We evaluate our distilled model on various downstream tasks in both domains. For speech tasks, we follow the SUPERB protocol \cite{superb}, and for music tasks, we use the MARBLE benchmark\cite{marble}. Table~\ref{tab:downstream_tasks_metrics} presents the downstream tasks, along with their respective evaluation metrics, datasets, and task domains. We implement the MARBLE tasks using two simple downstream linear layers and a learnable weighted sum of upstream features, as in speech tasks. They employ the Adam optimizer with a learning rate of 1e-3 and a batch size of 64, following MARBLE's downstream architecture. Each SSL model remains frozen during evaluation, with a learnable weighted sum applied to aggregate its hidden representations. A classifier, trained with labeled data, is then used to perform downstream tasks.

\begin{table*}[t]
  \centering
  \caption{Performance of different feature translators (Trans). Highlights hybrid translator's best performance on both HB/WL + MR.}
  \setlength{\tabcolsep}{6pt}
  \renewcommand{\arraystretch}{1.15}
  \begin{tabular}{ ll *{9}{c}}
    \toprule
        \textbf{Teacher} & \textbf{Trans} 
      & \textbf{ASR} & \textbf{KS} & \textbf{IC} & \textbf{ER} 
      & \textbf{SID} & \textbf{SingID} & \textbf{VID} & \textbf{PID} & \cellcolor{gray!20} \textbf{Avg (w/o ASR)} \\
          \cmidrule(lr){3-3} \cmidrule(lr){4-11} \vspace{-10pt}\\
                     &               
      & \textbf{(WER\% $\downarrow$)} 
      & \multicolumn{8}{c}{\textbf{(Acc\% $\uparrow$)}} \\
    \midrule
    HB + MR & Linear & 15.31 & 94.19        & 93.00 & 60.72       & 58.67   & 84.56           & 70.52        & 89.36        & \cellcolor{gray!20} 78.39      \\
    HB + MR & Hybrid & 15.29 & 94.97 & 91.96         & 61.98 & 64.38 & 87.26   & 71.75 & 89.82        & \cellcolor{gray!20} \textbf{80.30} \\
    HB + MR & Conv   & 15.32 & 93.48        & 85.95         & 62.72       & 64.27   & 86.48           & 69.21        & 90.45 & \cellcolor{gray!20} 78.93      \\
    \midrule
    WL + MR & Linear & 15.40 & 94.64        & 87.82         & 61.19 & 56.90   & 86.90           & 68.16        & 88.25        & \cellcolor{gray!20} 77.69      \\
    WL + MR & Hybrid & 14.53 & 94.87 & 93.83 & 59.21       & 56.88   & 84.34           & 68.33        & 88.77        & \cellcolor{gray!20} \textbf{78.03} \\
    WL + MR & Conv   & 14.82 & 93.93        & 86.66         & 58.98       & 57.30 & 84.84   & 70.61 & 89.40 & \cellcolor{gray!20} 77.39      \\
    \bottomrule
  \end{tabular}
\label{tab:translator}
\end{table*}


\begin{table*}[t]
  \centering
  \caption{Performance of different music weights ($w_M$) on MERT (with $w_S=1$). We select $\boldsymbol{w_M=0.3}$ as the best trade‐off.}
  \label{tab:weight}
  \setlength{\tabcolsep}{6pt}
  \renewcommand{\arraystretch}{1.15}
  \begin{tabular}{ ll *{9}{c}}
    \toprule
        \textbf{Teacher} & $\bm{w_M}$ 
      & \textbf{ASR} & \textbf{KS} & \textbf{IC} & \textbf{ER} 
      & \textbf{SID} & \textbf{SingID} & \textbf{VID} & \textbf{PID} & \cellcolor{gray!20} \textbf{Avg (w/o ASR)} \\
          \cmidrule(lr){3-3} \cmidrule(lr){4-11} \vspace{-10pt}\\
                     &               
      & \textbf{(WER\% $\downarrow$)} 
      & \multicolumn{8}{c}{\textbf{(Acc\% $\uparrow$)}} \\
    \midrule
    HB + MR & 0.1 & 14.29 & 94.87        & 94.73 & 62.62       & 68.95   & 81.28           & 67.81        & 81.84        & \cellcolor{gray!20} 78.87      \\
    HB + MR & 0.3 & 14.85 & 95.10 & 92.70         & 62.77 & 67.22           & 85.41   & 69.82 & 88.31        & \bfseries \cellcolor{gray!20} 80.19 \\
    HB + MR & 0.5 & 14.95 & 94.19        & 93.12         & 61.22       & 63.25           & 84.84   & 68.25        & 89.28        & \cellcolor{gray!20} 79.16      \\
    HB + MR & 1.0 & 15.31  & 94.19        & 93.00         & 60.72       & 58.67           & 84.56           & 70.52        & 89.36 & \cellcolor{gray!20} 78.39      \\
    \midrule
    WL + MR & 0.1 & 14.00 & 95.94 & 92.80         & 60.13       & 59.70   & 79.64           & 65.44        & 79.08        & \cellcolor{gray!20} 76.10      \\
    WL + MR & 0.3 &  14.19 & 94.84        & 95.40 & 60.67 & 58.90           & 81.85           & 70.96 & 85.47        & \bfseries \cellcolor{gray!20} 78.30 \\
    WL + MR & 0.5 & 14.37  & 95.42        & 93.65         & 59.42       & 58.21           & 82.00           & 70.07        & 86.99        & \cellcolor{gray!20} 77.97      \\
    WL + MR & 1.0 & 15.40 & 94.64        & 87.82         & 61.19       & 56.90           & 86.90   & 68.16        & 88.26 & \cellcolor{gray!20} 77.69      \\
    \bottomrule
  \end{tabular}
\end{table*}

\section{Results}
\label{sec:results}
\subsection{Overall Performance Comparison}

To demonstrate the effectiveness of our strategies, Table~\ref{tab:overall} presents the original teacher models, their respective distilled models, and the ensemble baseline. The ensemble model concatenates the representations of distilled speech and music models along their feature dimensions. From our ablation studies, we adopt Data-Domain Separation, a hybrid feature translator, and a music loss weight \(w_M\) = 0.3, speech loss weight \(w_S\) = 1 for both HuBERT and WavLM distillations with MERT.
Compared to their teacher models' distilled versions, our models significantly outperform them on cross-domain tasks: music tasks for HuBERT and WavLM, and speech tasks for MERT, while maintaining strong performance on their original tasks, resulting in a much more generalized model that excels across domains. Furthermore, when averaging performance excluding the ASR task, our models achieve 80.7\% for HB + MR and 77.65\% for WL + MR, both notably higher than any of the distilled teacher models. Additionally, Table~\ref{tab:overall} shows that both of our models achieve performance comparable to their ensemble baselines on most tasks, which concatenate the distilled teacher's representations along the feature dimension. Our HB + MR model experiences merely a slight drop of 0.8\% (80.40\% vs. 81.16\%) average accuracy, which suggests that our strategies not only minimize conflicts between teacher models but also reduce the number of parameters by 50\%(from 46M to 23M). 

\subsection{Ablations}

\subsubsection{\textbf{Data-Domain Separation}}

In multi-teacher distillation, different teacher models specialize in distinct domains, making domain-specific data assignment crucial for effective training. Since HuBERT and WavLM are pretrained on LibriSpeech and MERT on Music4All, we hypothesize that leveraging domain expertise enhances both training efficiency and task performance. 
To explore this, we evaluate a variant called Data-Domain Separation (DDS), where losses are computed exclusively from the speech teacher for speech data and from the music teacher for music data. This targeted approach ensures that each domain benefits from the most relevant teacher supervision.

To thoroughly investigate the impact of domain separation, we conduct three experiments. First, we use DDS. Second, we train on all data (LibriSpeech + Music4ALL) without distinguishing domains. Third, we train using only LibriSpeech. We apply these strategies to two model combinations—HuBERT + MERT and WavLM + MERT—to verify whether the method remains consistent across different teacher model pairings.

Table~\ref{tab:data_domain} underscores the importance of considering data domains when distilling a model. First, by comparing the model trained on all data without Data-Domain Separation (DDS) and the one trained on LibriSpeech, we observe a substantial improvement in music-related tasks, and have a higher overall average (78.68\% vs. 77.23\%) and (77.29\% vs. 76.99\%) for HB/WL + MR, respectively. This suggests the importance of integrating music data into the distillation. 

Moreover, when comparing the DDS model with the one without DDS, we observe a significant drop in the IC task performance for the latter. This result suggests that mixing domains introduces conflicting representations, leading to a notable decline in IC performance. The DDS model also achieves a higher average score in both model combinations (78.71\% vs. 78.68\%) and (77.69\% vs. 77.29\%), indicating that it serves as a more generalized model. While separating data domains slightly lowers music task performance, it prevents a steep decline in speech-oriented tasks. Furthermore, our subsequent experiments have identified methods to recover music-related performance losses, ensuring that domain separation remains the preferred strategy overall.
\\

\subsubsection{\textbf{Feature Translator}}

Teacher models from different domains produce distinct features, making direct alignment with a simple linear prediction head insufficient. While a linear translator captures global shifts between domains, it struggles with localized differences, whereas a large transformer-based translator can model complexities but risks overfitting and harming generalization. 

To address these challenges, we evaluate three types of translators. The \textbf{Linear Translator} uses a single fully connected layer to project teacher embeddings, providing a straightforward global alignment. The \textbf{Convolution Translator} employs convolutional layers to capture localized patterns and handle non-linear feature differences. Table~\ref{tab:translator} shows that using a fully convolutional translator benefits tasks such as VID, SingID, PID, and SID for both HuBERT and WavLM. However, this approach can degrade performance on speech-centric tasks like KS and IC, suggesting it may overlook semantic information for speech.

In response, we adopt a \textbf{Hybrid Translator} that applies a linear translator to the speech teacher and a convolutional translator to the music teacher, thereby balancing both global and local adaptations. The hybrid approach leverages the strengths of both translators: it preserves delicate speech features through linear mapping while effectively adapting complex music features via convolutional processing. This design not only resolves the observed degradation in IC performance under a fully convolutional translator (e.g., improving IC from 85.95\% to 91.96\% with the HuBERT teacher) but also maintains robust performance across other tasks. In particular, the hybrid model consistently achieves the best ASR performance across two teacher combinations, further validating its effectiveness in preserving critical speech information during distillation. Our results demonstrate that the hybrid translator offers an effective solution for multi-domain feature alignment, with the highest average accuracy (80.3\% and 78.03\%), reducing conflicts between speech and music representations and optimizing the overall distillation quality.

\subsubsection{\textbf{Loss Weight Adjustment}}

Our experiment found MERT significantly easier to distill than the other two models. Therefore, equal loss weighting can cause easier models like MERT to dominate training while more complex models like HuBERT and WavLM struggle, leading to suboptimal performance. To address this, we aim to find an optimal weighting ratio that minimizes gradient conflicts and ensures balanced knowledge transfer for speech and music tasks. We keep the speech loss weight \(w_S\) fixed at 1 and vary \(w_M\) across \{0.1, 0.3, 0.5, 1\} to evaluate its impact on learning balance. The total loss is computed as:  
\begin{equation}
    \mathcal{L}_{\text{total}} = \mathcal{L}_{\text{speech}} + w_M \cdot \mathcal{L}_{\text{music}}
\end{equation}

Table~\ref{tab:weight} illustrates the impact of loss weighting on performance. When using HuBERT as the speech teacher, assigning a higher weight (\(w_M = 1\)) improves music task performance. For instance, PID accuracy rises from 81.84\% at \(w_M = 0.1\) to 89.36\% at \(w_M = 1\), but at the cost of degraded speech task performance, with SID accuracy dropping from 68.95\% to 58.67\%. To find the best balance, we select the weight that yields the highest overall average, indicating the most well-rounded model. From Table~\ref{tab:weight}, \textbf{\(w_M = 0.3\)} achieves the best trade-off with the highest overall performance (80.36\%). Similarly, when using WavLM as the speech teacher, \textbf{\(w_M = 0.3\)} also provides the best balance, achieving an average performance of 77.95\%.  


\subsection{Few-Shot Learning}

Real-world applications frequently face situations where only a handful of labeled examples are available, and it is crucial to assess how well a unified model can adapt under such low-resource conditions. 
To investigate this, we evaluate the few-shot learning performance of our models, which adopt DDS and a hybrid translator, on two tasks from each domain: Keyword Spotting (KS), Emotion Recognition (ER) for the speech domain, and Vocal Identification (VID), Singer Identification (SingID) for the music domain. In each task, we randomly choose 20 shots per class to form a training set, and we repeat this process to create 5 different sets to ensure robustness. We then fine-tune the downstream classifier on each split and report the average accuracy. To ensure consistency, the test sets used in these few-shot evaluations are the same as those used in our earlier ablation studies.

Table~\ref{tab:fewshot} illustrates the robustness of our model in few-shot learning scenarios. We observe that the WL + MR model achieves the highest scores across all tasks, while the HB + MR model performs better on ER and SingID, with only a negligible drop ($<$0.5\%) on KS and VID. These results demonstrate that our unified model outperforms its distilled teacher models in few-shot learning, showcasing that our approach reduces domain bias and enhances the model's ability to handle both speech and music tasks effectively, even with limited data.


\begin{table}[t!]
  \centering
  \caption{Performance (Acc\% ↑) of distilled models on Few-Shot Learning. Along with the standard deviation of each score across five data splits. The \textbf{bold numbers} indicate that our model outperforms both of its distilled teacher models.}
  \label{tab:fewshot}
  \setlength{\tabcolsep}{6pt}
  \renewcommand{\arraystretch}{1.15}
  \begin{tabular}{ l c c c c c c }
    \toprule
        \textbf{Teacher} & \textbf{KS} & \textbf{ER} & \textbf{SingID} & \textbf{VID} & \cellcolor{gray!20} \textbf{Avg} \\
    \midrule
    HuBERT    & 61.20\scriptsize{±2.7} & 46.92\scriptsize{±3.2} & 43.01\scriptsize{±1.7} & 51.08\scriptsize{±1.4} & \cellcolor{gray!20} 50.55 \\
    WavLM     & 68.60\scriptsize{±2.1} & 42.74\scriptsize{±2.2} & 43.04\scriptsize{±1.6} & 49.78\scriptsize{±1.9} & \cellcolor{gray!20} 51.04 \\
    MERT      & 36.13\scriptsize{±2.2} & 41.26\scriptsize{±3.0} & 48.60\scriptsize{±0.6} & 62.49\scriptsize{±2.7} & \cellcolor{gray!20} 47.12 \\
    HB+MR   & 61.04\scriptsize{±2.5} & \textbf{47.39}\scriptsize{±3.3} & \textbf{50.96}\scriptsize{±2.2} & 62.16\scriptsize{±1.9} & \cellcolor{gray!20} \textbf{55.39} \\
    WL+MR    & \textbf{69.77}\scriptsize{±2.8} & \textbf{43.05}\scriptsize{±2.6} & \textbf{49.70}\scriptsize{±1.0} & \textbf{62.84}\scriptsize{±2.3} & \cellcolor{gray!20} \textbf{56.34} \\
    \bottomrule
  \end{tabular}
  \vspace{-5pt}
\end{table}
\vspace{-5pt}

\section{Conclusion}

This paper presented a multi-teacher distillation framework that unifies domain-specific speech and music representation models into a single, comprehensive audio model. Our approach leverages domain-adaptive strategies, including Data-Domain Separation, loss weight adjustment, and feature translators to bridge the gap between heterogeneous teacher models. Extensive experiments across diverse downstream tasks demonstrate that our unified model generalizes well to speech and music applications and outperforms distilled baselines while reducing computational complexity. Additionally, our few-shot learning experiments highlight the model's robustness in low-resource scenarios and underscore its practicality in real-world applications. This work lays the foundation for efficient multi-domain audio processing, with potential applications in areas like speech-based music recommendation, multimedia indexing, and more. Future research will further refine our methods and extend the framework to incorporate additional audio domains, such as environmental sounds or multi-speaker interactions, ultimately advancing the development of universal audio models.

\section{Limitations}

While our proposed framework yields robust performance across multiple tasks, several areas warrant further improvement. First, the current approach requires manual tuning of the loss weight ratio between speech and music models. Although careful tuning has led to a balanced performance, developing an automated or adaptive weighting scheme could streamline this process and enhance training efficiency. Additionally, although the unified model performs competitively against domain-specific models, a slight performance gap remains on certain tasks. Future work will explore advanced strategies to preserve domain-specific details while maintaining overall generalization. Finally, while our experiments validate the model on standard benchmarks, further evaluation on real-world applications that simultaneously exploit speech and music representations would provide additional practical insights.

\bibliographystyle{IEEEtran}
\bibliography{mybib}

\end{document}